\documentclass[twocolumn]{aastex63}
\usepackage{graphicx}
\usepackage{amsmath}
\usepackage{enumitem}
\usepackage[normalem]{ulem}

\shorttitle{Merger Timescale of subparsec SMBHBs}
\shortauthors{Nguyen et al.}

\begin{document}
\def\redtext#1{#1}
\def\redtext#1{{\color{red}#1}}
\def\bluetext#1{#1}
\def\bluetext#1{{\color{blue}#1}}
\def\greentext#1{#1}
\def\greentext#1{{\color{green}#1}}

\title{Pulsar Timing Array Constraints on the Merger Timescale of Subparsec Supermassive Black Hole Binary Candidates}

\correspondingauthor{Tamara Bogdanovi\'c}
\email{tamarab@gatech.edu, khainguyen@gatech.edu}

\author[0000-0003-3792-7494]{Khai Nguyen}
\affiliation{Center for Relativistic Astrophysics, School of Physics, Georgia Institute of Technology, Atlanta, GA 30332, USA}

\author[0000-0002-7835-7814]{Tamara Bogdanovi\'c}
\affiliation{Center for Relativistic Astrophysics, School of Physics, Georgia Institute of Technology, Atlanta, GA 30332, USA}

\author[0000-0001-8557-2822]{Jessie C. Runnoe}
\affiliation{Department of Physics \& Astronomy, Vanderbilt University, 6301 Stevenson Center Ln, Nashville, TN 37235, USA}

\author[0000-0001-8217-1599]{Stephen R. Taylor}
\affiliation{Department of Physics \& Astronomy, Vanderbilt University, 6301 Stevenson Center Ln, Nashville, TN 37235, USA}

\author[0000-0003-4961-1606]{Alberto Sesana}
\affiliation{Dipartimento di Fisica ``G. Occhialini", Universit\`{a} degli Studi di Milano-Bicocca, Piazza della Scienza 3, 20126 Milano, Italy}

\author[0000-0002-3719-940X]{Michael Eracleous}
\affiliation{Department of Astronomy \& Astrophysics and Institute for Gravitation and the Cosmos, Pennsylvania State University\\
525 Davey Lab, University Park, PA 16802}

\author[0000-0002-8187-1144]{Steinn Sigurdsson}
\affiliation{Department of Astronomy \& Astrophysics and Institute for Gravitation and the Cosmos, Pennsylvania State University\\
525 Davey Lab, University Park, PA 16802}

\begin{abstract}
We estimate the merger timescale of spectroscopically-selected, subparsec supermassive black hole binary (SMBHB) candidates by comparing their expected contribution to the gravitational wave background (GWB) with the sensitivity of current pulsar timing array (PTA) experiments and in particular, with the latest upper limit placed by the North American Nanohertz Observatory for Gravitational Waves (NANOGrav). We find that the average timescale to coalescence of such SMBHBs is $\langle t_{\rm evol} \rangle > 6\times 10^4\,$yr, assuming that their orbital evolution in the PTA frequency band is driven by emission of gravitational waves. If some fraction of SMBHBs do not reside in spectroscopically detected active galaxies, and their incidence in active and inactive galaxies is similar, then the merger timescale could be $\sim 10$ times longer, $\langle t_{\rm evol} \rangle > 6\times 10^5\,$yr. These limits are consistent with the range of timescales predicted by theoretical models and imply that all the SMBHB candidates in our spectroscopic sample could be binaries without violating the observational constraints on the GWB. This result illustrates the power of the multi-messenger approach, facilitated by the PTAs, in providing an independent statistical test of the nature of SMBHB candidates discovered in electromagnetic searches.
\end{abstract}



\section{Introduction}\label{sec:intro}

Over the past decade spectroscopic searches have identified about a hundred supermassive black hole binary (SMBHB) candidates at subparsec orbital separations \citep{bon12,bon16,eracleous12,decarli13,ju13,shen13,liu14,li16,runnoe15,runnoe17,wang17,guo19}. These searches rely on detection and long term monitoring of the Doppler shift in the optical emission-line spectrum of active galactic nuclei (AGNs), that arise as a consequence of SMBHB orbital motion, under assumption that at least one of its constituent supermassive black holes (SMBHs) can shine as an AGN \citep{bbr80, gaskell83, gaskell96}. 

With a cadence of observations anywhere from days to years, spectroscopic searches are in principle sensitive to binaries with orbital periods in the range $\sim 10-100{\rm s}$ years and separations of at most ${\rm few} \times 10^4 r_g$ \citep[$r_g= G M/c^2$ and $M$ is the binary mass;][]{pflueger18}. 
For each observed SMBHB with mass $10^8 M_{\odot}$, a comparable mass ratio, and orbital separation of about $10^4 r_g$, the projection factors (i.e., orientation of the binary orbit relative to the observer's line of sight) imply a few undetected binaries, and possibly more if some fraction of SMBHBs do not exhibit AGN signatures. Furthermore, for every SMBHB in the ``detectable" range, there should be over 200 more gravitationally bound systems with similar properties but at larger separations, where they cannot be detected by optical spectroscopic searches \citep{pflueger18}. Thus, any SMBHB detected using this technique would represent the tip of the iceberg of binaries that escape detection because they are either: (a) under-luminous, (b) have unfavorable orientation, (c) have orbital velocities that are too low or (d) reside in a portion of the sky not covered by the search \citep[see][for a systematic study of these effects]{kelley20}.

The main complication of spectroscopic searches is the fact that the velocity-shift and modulation of emission lines around their rest frame wavelength is not unique to SMBHBs \citep[e.g.,][]{eracleous12,popovic12,barth15, guo19}, making it difficult to uniquely identify binaries. This is of importance because if any of detected SMBHB candidates are real binaries, they are direct progenitors of systems that coalesce due to the emission of gravitational waves (GWs). More specifically, they imply some number of SMBHBs inspiraling toward coalescence, whose GW signal is reaching Earth at this very moment. If there were many of them, the stochastic superposition of their GWs would have already been detected by the pulsar timing arrays (PTAs).

PTAs seek to detect GWs by searching for correlations in the timing observations of a network of millisecond pulsars. Currently, there are three such experiments in operation: the North American Observatory for Gravitational Waves \citep[NANOGrav;][]{mclaughlin13}, the European PTA \citep[EPTA;][]{desvignes16}, and the Parkes PTA \citep[PPTA;][]{hobbs13}. Together they form the International PTA \citep[IPTA;][]{verbiest16}.  At this time, PTA searches for an isotropic stochastic GW background (GWB) are starting to reach sensitivities necessary to probe backgrounds of astrophysical origin \citep{shannon15, lentati15, arzoumanian18}. 

The massive ($M > 10^8 M_{\odot}$) and nearby ($z \approx 1- 2$) SMBHBs are the major contributors to low frequency GWs sought by PTAs \citep{sesana08}. Although current limits are still insufficient to place stringent constrains on the cosmic population of SMBHBs \citep{middleton18}, they can be used to test candidates assembled from electromagnetic observations. For example, \citet{sesana18} found that the GWB implied by a sample of $\sim 150$ photometrically-selected SMBHB candidates (based on potential periodicity in their light curves) is in tension with the current most stringent PTA upper limits, implying that at least some fraction are false positives. A similar technique was used to place limits on the presence of SMBHBs in periodic blazars \citep{holgado18} and in ultraluminous infrared galaxies \citep{inayoshi18}.

In this work, we use a spectroscopic sample of SMBHB candidates from \citet[][hereafter E12]{eracleous12}, who searched for $z < 0.7$ Sloan Digital Sky Survey quasars \citep[DR7;][]{schneider10}, with broad H$\beta$ lines offset from the rest frame of the host galaxy by $\gtrsim {\rm few} \times 100\,{\rm km\,s^{-1}}$. Based on this criterion, E12 selected 88 SMBHB candidates for observational follow-up from an initial group of about 15,900 objects. From this sample of candidates we infer the underlying population of binaries that are inspiraling due to the emission of GWs. Instead of taking a forward modeling approach to this problem common in the literature, in which we would adopt a particular model for orbital evolution from the subparsec scales to the GW band, we ask: If the GWB signal of the sample of hypothetical SMBHBs that we consider is to be smaller than the sensitivity limit of the PTAs, what is the lower limit on their evolution timescale?

\section{Methods}\label{sec:methods}

\begin{figure*}[t]
\centering
\includegraphics[width=1.0\textwidth, clip=true]{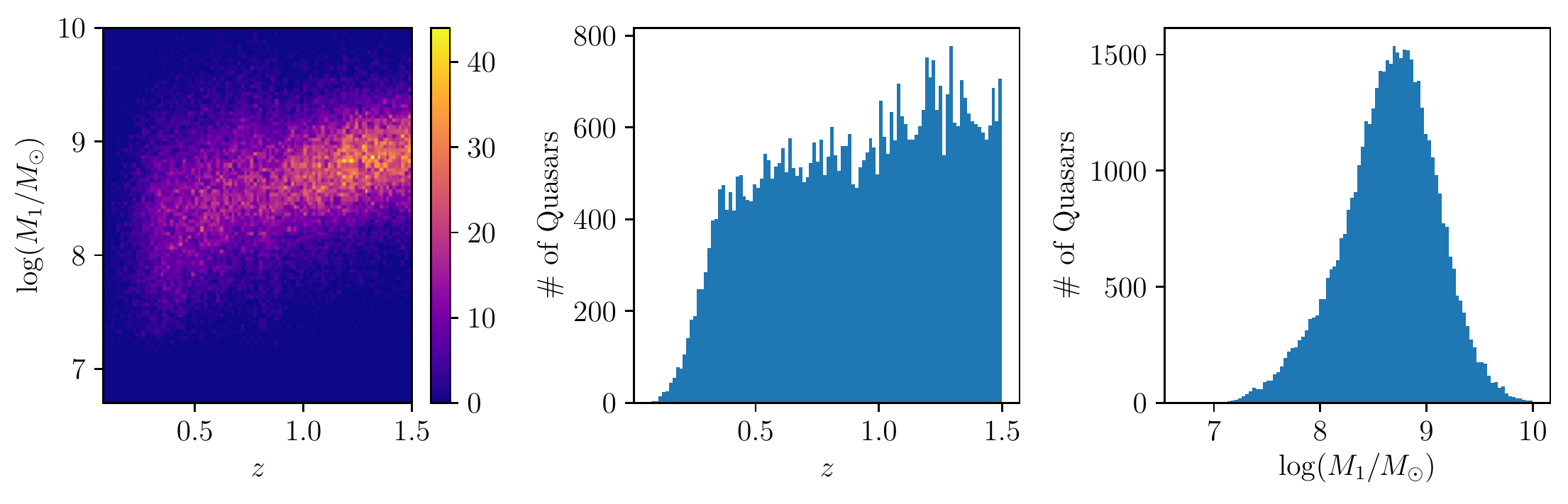}
\caption{Distribution of SMBHs in mass and redshift inferred from the SDSS DR7 quasar catalog (left pannel) and histograms of the distribution in redshift (middle) and mass (right). The one-dimensional distributions are the projections of the two-dimensional distribution on the mass and redshift axes. We assume that the mass distribution of primary SMBHs in hypothesized binaries has the same shape as for the SMBHs in SDSS quasars. The colorbar marks the number of quasars.}
\label{fig:Qcatalog}
\end{figure*}  

\subsection{Merger Rate of SMBHBs}\label{sec:mRate}

In order to determine the GWB contributed by a population of SMBHBs, we calculate their differential merger rate 
\begin{equation}
\label{eqn:mrate}
\frac{d^5 N}{dM_1 \, d \tilde{a}\, dq \, dz \, dt_r} = \frac{\nu (M_1;z)}{t_{\rm evol} (M_1, \tilde{a}, q)} \, \frac{\rho(\tilde{a},q,z)}{P_{\rm bias}} \, , 
\end{equation}
where $M = M_1 + M_2$ is the binary mass, $q = M_2/M_1 < 1$ is the mass ratio with $M_1$ ($M_2$) being the mass of the primary (secondary) SMBH, $\tilde{a}\equiv a/r_g$ is the dimensionless semimajor axis, $z$ is redshift, and $t_r$ is time measured in the rest frame of the SMBHB. 

The quantities on the right hand side of equation~\ref{eqn:mrate} represent the distribution of SMBHB properties inferred from the E12 sample of candidates by correcting for selection effects. Here, $\nu(M_1;z)$ is the mass distribution as a function of $z$ of spectroscopically detectable SMBHBs (see \S\ref{sec:massDistribution}). 
The parameter $t_{\rm evol} (M_1, \tilde{a}, q)$ is the timescale for evolution of a SMBHB from a separation at which it was detected ($\sim 10^4 r_g$ for spectroscopically targeted binaries) to coalescence.  It is usually estimated from the merger rate as $dN/dt_r \approx N/t_{\rm evol}$ and it depends on the SMBHB parameters, as well as the physical mechanisms that drive binary to coalescence \citep[gas, stellar torques and GW emission;][]{sesana13}. The function $\rho(\tilde{a},q,z)$ is the probability distribution of SMBHB candidates given $\tilde{a}$, $q$ and $z$, introduced in \S\ref{sec:selectionbias}. $P_{\rm bias}$ is a probability that a SMBHB is detected by the E12 spectroscopic search given the selection effects inherent to this technique (see \S\ref{sec:Pbias}).

\subsection{Distribution of SMBHBs -- $\nu(M_1;z)$ and $\rho(\tilde{a},q,z)$}\label{sec:massDistribution}\label{sec:selectionbias}

We derive the mass distribution of primary SMBHs in all hypothesized binaries within the redshift range $0 < z < 1.5$\footnote{This expression implies an upper limit in redshift that encloses most of the GWB from SMBHBs detected by PTAs. We justify this assumption in \S~\ref{sec:results}.} by assuming that it has the same shape as the mass distribution of SMBHs powering SDSS quasars but a different normalization, since only a small fraction of quasars may host binaries. This is reasonable since we expect that the primary SMBHs in the E12 sample would have formed in the same way as the rest of the SDSS quasars powered by isolated SMBHs: through prior mergers and accretion. Thus, we describe the primary SMBHs using the mass distribution of the quasars from the SDSS DR7 catalog. These masses are obtained using the virial SMBH mass estimators, based on the continuum luminosities and the H$\beta$ or \ion{Mg}{2} lines \citep{shen11}. We use measurements for which the observed line profiles are fit with reduced chi-squared between about 0.8 and 1.5, ensuring a reliable fit, and eliminate quasars with broad absorption lines, which may have inaccurate mass estimates. 

The left panel of Figure~\ref{fig:Qcatalog} shows the resulting SMBH mass distribution for the SDSS quasars. This is a distribution whose normalization evolves with redshift (middle panel), with a majority of SMBH masses in the range $10^{7-10}\,M_\odot$ and a median of $\sim 5 \times 10^8\,M_\odot$ (right panel).  It is worth noting that because SDSS is a flux limited survey, at every redshift there are active galaxies that are below its detection threshold. This is reflected in a dearth of SMBHs with masses $\lesssim 10^8\,M_\odot$ beyond $z\approx 0.5$ in the left panel of Figure~\ref{fig:Qcatalog}. This is of interest because if some fraction of these objects are tracers of inspiraling SMBHBs, they represent a contribution to the GWB that is unaccounted for. We examine the impact of this selection effect on the resulting GWB in \S\ref{sec:results}.

The virial SMBH mass measurements, like the ones obtained from the SDSS DR7 catalog, are known to be subject to Malmquist bias \citep{shen08}. This effect arises because the underlying SMBH mass distribution in the mass range of interest is bottom heavy (i.e., there are more SMBHs toward lower masses), and as a result more objects scatter from the low-mass bins to high than the other way around. Thus, the observed virial mass distribution for the SDSS sample is biased high by about 0.55 dex relative to the ``true" underlying distribution. We evaluate the impact of this effect by performing calculations of the merger rate with $\nu(M_1;z)$  (a) uncorrected for Malmquist bias, as shown in Figure~\ref{fig:Qcatalog}, and (b) corrected for this bias by shifting the distribution to lower masses by 0.55 dex. The median SMBH mass of the corrected distribution is then about $10^8\,M_\odot$.

Finally, we obtain the normalization of the SMBHB mass distribution in either scenario by scaling down the SMBH mass distribution function in Figure~\ref{fig:Qcatalog} in such way, that in the redshift range $0 < z < 0.7$ the number of objects corresponds to 88, the number of SMBHB candidates in the E12 sample. The resulting number of SMBHBs out to $z=1.5$ inferred in this way is 285 (see however the discussion of selection effects in \S\ref{sec:Pbias}).

\begin{figure*}[t]
\centering
\includegraphics[width=1.0\textwidth, clip=true]{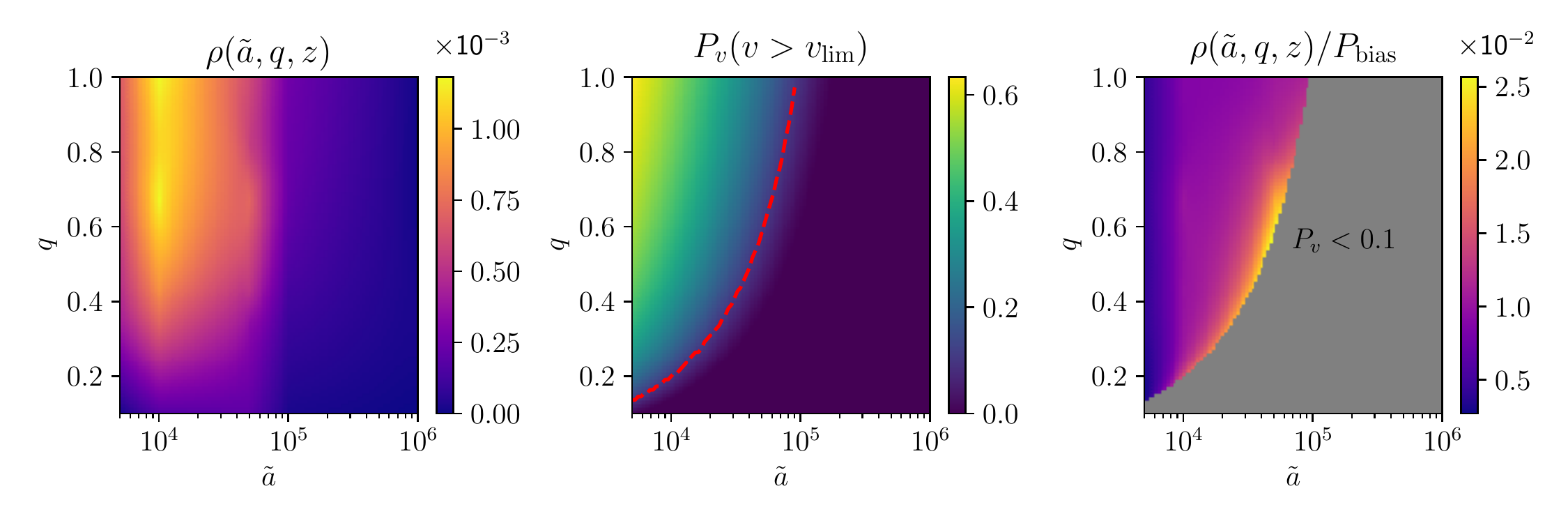}
\caption{{\it Left panel:} Probability density distribution of the SMBHB candidates from the E12 sample, $\rho(\tilde{a},q,z)$, integrated over redshift. {\it Middle:} Probability of detecting a SMBHB with radial component of orbital velocity greater than $v_{\rm lim} = 350\, \textrm{km s}^{-1}$. The red dashed line marks $P_v = 0.1$ contour. {\it Right:} The probability density distribution of the inferred SMBHBs population after accounting for the selection effects, $\rho(\tilde{a},q,z)/P_{\rm bias}$. The probability density in the greyed-out region is set to zero (see \S\ref{sec:Pbias}).}
\label{fig:Pv}
\end{figure*}  

In order to aid the interpretation of spectroscopic SMBHB candidates, \citet{nguyen16} and \citet{nguyen19} developed a semi-analytic model to calculate the broad emission-line profiles emitted from circumbinary accretion flows associated with subparsec SMBHBs. They found that the modeled profiles show distinct statistical properties as a function of the binary semimajor axis and mass ratio and that as a result, broad emission lines can be used to infer their distribution. A subsequent analysis presented in \citet{nguyen20} showed that as a population, the E12 SMBHB candidates favor an average value of the semimajor axis corresponding to $\log \tilde{a} \approx 4.20$ with standard deviation of 0.42, and comparable mass ratios, $q>0.5$.

The left panel of Figure~\ref{fig:Pv} shows the resulting probability density distribution for the E12 sample of SMBHB candidates from \citet{nguyen20}, $\rho(\tilde{a},q,z)$, integrated over redshift. The distribution shown in the figure is normalized in such way that when integrated with respect to $\tilde{a}$, $q$ and $z$ returns 88, the total number of the E12 SMBHB candidates. In the absence of other information about the properties of the SMBHB candidates with redshift $z\geq 0.7$, we assume that they are characterized by the same distribution, $\rho(\tilde{a},q,z)$, as the E12 sample. 

One can show that each hypothetical SMBHB, characterized by the distribution of $M_1$, $\tilde{a}$ and $q$ described here, is more likely to have an evolution time scale longer than a Hubble time, if its evolution was driven solely by the emission of GWs. Therefore, these SMBHBs can evolve into the PTA frequency band only if their evolution at larger separations is driven by gas and / or stars.

\subsection{Probability of detection -- $P_{\rm bias}$}\label{sec:Pbias}

If all objects in the E12 sample are true binaries, one would expect an underlying population larger than 88, given the selection effects of the search. We account for two such effects: one is a probability of detection given a partial sky coverage of the SDSS DR7 spectroscopic survey, which corresponds to $P_{\rm sdss}\approx 1/4$. The other is a probability, $P_v$, that a SMBHB has the radial component of orbital velocity greater than some threshold value that defines the sensitivity of the search, $v > v_{\rm lim}$. Note that the latter probability accounts for the fact that some fraction of SMBHBs escape detection either because they have unfavorable orientation or because their orbital velocity is lower than $v_{\rm lim}$ regardless of orientation, as mentioned in \S\ref{sec:intro}. The total probability of detection is then $P_{\rm bias}=P_{\rm sdss}\, P_v$. Note that so far we do not account for the fact that some unknown fraction of SMBHBs may reside in systems that do not exhibit AGN signatures (see discussion in \S~\ref{sec:discussion}).

Assuming for simplicity SMBHBs on circular orbits and that the measured radial velocity traces the motion of the primary SMBH, $P_v$ can be expressed analytically \citep{pflueger18}
\begin{equation}
P_v(v>v_{\rm lim})=1-\frac{2}{\pi}\left(\arcsin \zeta + \zeta \ln \left[ \frac{1+\cos(\arcsin \zeta)}{\zeta}     \right]         \right)  
\label{eqn:Pv}
\end{equation}
where $\zeta =  \tilde{a}^{1/2} (1/q+1) (v_{\rm lim}/c)$ is a dimensionless parameter. Note that the premise that $v$ is associated with the primary SMBH marks a departure from that commonly adopted by spectroscopic searches, which assume that $v$ associated with the secondary instead. This is supported by modeling, that indicates that in most SMBHB configurations the accretion disk around the primary makes the dominant contribution to the H$\beta$ broad emission-line flux \citep{nguyen20}. See \S~\ref{sec:results} for a description of how our results are affected by this assumption.

The middle panel of Figure~\ref{fig:Pv} shows $P_v$ calculated for $v_{\rm lim} = 350\,\textrm{km s}^{-1}$. This value corresponds to the smallest velocity offset measured in the E12 sample, in the first epoch of observations, and is representative of the sensitivity achieved by the search. Figure~\ref{fig:Pv} illustrates that the probability of detection increases with $q$, as the orbital speed of the primary SMBH becomes more pronounced. Similarly,  $P_v$ decreases with $a$, as the binary orbital velocity decreases with separation.

To derive the probability density of the underlying SMBHB population, and factor out the selection effects described above, we divide $\rho(\tilde{a},q,z)$ with $P_{\rm bias}$ and show the result in the right panel of Figure~\ref{fig:Pv}. This distribution indicates an increasing number of SMBHBs at larger orbital separations, as expected if wider binaries are evolving more slowly. This approach however cannot be used to reliably extrapolate the number of SMBHBs in the region where the sensitivity of the search drops significantly. This region is marked by dark blue colors in the left and middle panels of Figure~\ref{fig:Pv}   and is outlined by the red dashed line in the middle panel with a $P_v = 0.1$ contour. In order to mitigate the uncertainty caused by small number statistics we set $\rho(\tilde{a},q,z) = 0$  where $P_v < 0.1$ and make no predictions for the underlying SMBHB population in the greyed out area in the right panel of Figure~\ref{fig:Pv}. We account for the effect of truncation in $\rho(\tilde{a},q,z)$ by rescaling its normalization to ensure that when integrated in terms of $\tilde{a}$, $q$ and $z$ it still returns 88.

The inferred number of SMBHBs with $z< 0.7$ obtained in this way, calculated by integrating the distribution $\rho(\tilde{a},q,z) / P_{\rm bias}$ shown in the right panel, is around 1492, indicating that for every SMBHB detected in this parameter space there are about 16 more that escape detection on average, because of selection effects. Extending the same reasoning to SMBHBs with $z< 1.5$ implies about $285\times17 = 4845$ binaries in this redshift range. 

It is worth mentioning that an additional selection effect introduced by spectroscopic searches is a probability that a SMBHB has a change in radial velocity, measured as an epoch-to-epoch modulation in the velocity offset of the broad emission lines, larger than some threshold value, $\Delta v > \Delta v_{\rm lim}$ \citep[see][]{pflueger18}. We neglect this effect as it was not used to eliminate any SMBHBs in the E12 sample thus far.

\subsection{Calculation of the Gravitational Wave Background}\label{sec:GWBstrain}

%

We calculate the GWB strain as a function of the observed frequency, $h_c(f)$, following the approach described in \citet{phinney01} and \citet{sesana04,sesana08} 
\begin{equation}
\label{eqn:phinney}
h_c^2(f)=\frac{4G}{\pi c^2 f^2}\int dz \, \frac{n(z)}{1+z}\frac{dE_{\rm GW}}{d \ln f_r} \,,
\end{equation}
where $f_r=f(1+z)$ is the GW frequency in the rest frame of the binary. $E_{GW}$ is the energy emitted in GW, which for a circular SMBHB can be expressed as
\begin{equation}
\frac{dE_{\rm GW}}{d \ln f_r}=\frac{\pi^{2/3}}{3G}(G\mathcal{M})^{5/3}f_r^{2/3} \,,
\label{eq_Egw}
\end{equation}
and $\mathcal{M} = M_1\,q^{3/5}/(1+q)^{1/5}$ is the chirp mass. In this calculation, equation~\ref{eq_Egw} represents SMBHBs emitting in the frequency band of NANOGrav that evolve primarily due to the emission of GWs, as opposed to gas and stellar torques. We discuss the implications of this assumption in \S\ref{sec:discussion}. $n(z)$ represents number of binary mergers per unit comoving volume per unit redshift, $n(z)=d^2N / dz\,dV_{\rm c}$, and thus
\begin{equation}
n(z)=\iiint dM_1 \, d\tilde{a} \,  dq \,  \frac{d^5 N}{dz \, dM_1 \, d \tilde{a} \, dq \, dt_r}  \frac{dt_{\rm r}}{dV_{\rm c}} \,,
\end{equation}
where the relationship between time and comoving volume is given by $dt_{\rm r}/dV_{\rm c} = [4\pi c\,(1+z)\,d_{M}^2(z)]^{-1}$. The comoving distance is given by
\begin{equation}
\label{eqn:comovingdist}
d_{M}(z)=\frac{c}{H_0} \int_0^z \frac{dz'}{\sqrt{\Omega_M(1+z')^3+\Omega_{\Lambda}}} \,,
\end{equation} 
where we assume a flat universe with $\Omega_M=0.315$, $\Omega_{\Lambda}=0.685$, $\Omega_k=0$, $H_0=67.4\, \text{km s}^{-1} \text{Mpc}^{-1}$ \citep{planck18}. Combining equations \ref{eqn:phinney} -- \ref{eqn:comovingdist} with equation~\ref{eqn:mrate} we obtain
\begin{multline}
\label{eqn:GWB}
h_c^2(f)=  \frac{G^{5/3}}{3\pi^{4/3}c^3} \frac{1}{f^{4/3}} \iiiint dM_1\, dz \, d\tilde{a}\, dq   \\
\frac{1}{t_{\rm evol}(M_1, \tilde{a}, q)} \, \frac{\nu(M_1;z)}{(1+z)^{4/3}}  \, \frac{\rho(\tilde{a},q,z)}{P_{\rm bias}} \frac{\mathcal{M}^{5/3}}{d_{M}^2(z)} \,,
\end{multline}
and subsequently,
\begin{multline}
\label{eqn:GWBlowerz}
h_c^2(f)=  \frac{G^{5/3}}{3\pi^{4/3}c^3}   \frac{1}{f^{4/3}} \frac{1}{\langle t_{\rm evol}\rangle} \iiiint dM_1\, dz \, d\tilde{a}\, dq   \\
\frac{\nu(M_1;z)}{(1+z)^{4/3}}  \, \frac{\rho(\tilde{a},q,z)}{P_{\rm bias}} \frac{\mathcal{M}^{5/3}}{d_{M}^2(z)} \,.
\end{multline}

\begin{deluxetable}{cccc}[t]
\tablecaption{GWB strain at $f=1{\rm yr}^{-1}$}\label{table:hc}
\tablehead{$\langle t_{\rm evol}\rangle$/yr & $h_{c1}$ & $h_{c2}$ & $h_{c3}$}
\startdata
$10^9$ & $2.80\times10^{-17}$ & $3.24\times 10^{-17}$ & $1.13\times 10^{-17}$  \\
$10^8$ & $8.85\times 10^{-17}$ & $1.02\times 10^{-16}$ & $3.57\times 10^{-17}$ \\
$10^7$ & $2.80\times 10^{-16}$ & $3.24\times 10^{-16}$ & $1.13\times 10^{-16}$ \\
$10^6$ & $8.85\times 10^{-16}$ & $1.02\times 10^{-15}$ & $3.57\times 10^{-16}$ \\
$10^5$ & $2.80\times 10^{-15}$ & $3.24\times 10^{-15}$ & $1.13\times 10^{-15}$ \\
$10^4$ & $8.85\times 10^{-15}$ & $1.02\times 10^{-14}$ & $3.57\times 10^{-15}$ 
\enddata
\tablecomments{$\langle t_{\rm evol}\rangle$ -- merger timescale. $h_{c1}$, $h_{c2}$ -- GWB strain amplitudes for SMBHBs at $z<0.7$ and $z<1.5$, respectively, uncorrected for Malmquist bias. 
$h_{c3}$ -- GWB strain amplitude for SMBHBs at $z<1.5$, corrected for Malmquist bias. See \S\ref{sec:results} for more detail.}
\end{deluxetable}

Equating equations~\ref{eqn:GWB} and \ref{eqn:GWBlowerz} yields a definition of $\langle t_{\rm evol} \rangle$, a characteristic merger timescale for evolution of the ensemble of SMBHBs in the redshift range $0 < z < 1.5$, from the separations at which the E12 candidates are typically detected ($\sim 10^4\,r_g$) to coalescence. $\langle t_{\rm evol} \rangle$ is calculated as an average over the distributions in $M_1$, $z$, $\tilde{a}$ and $q$, and weighted by the factors in equation~\ref{eqn:phinney}, of which $dE_{\rm GW}/d \ln f_r$ puts weight on the loudest binaries in a given frequency interval.

 At such large initial separations the evolution of SMBHBs headed for coalescence is driven by stellar and gas torques. This allows us to decouple $\langle t_{\rm evol} \rangle$ from the calculation of the GW signal of such SMBHBs in the NANOGrav band, where we assume that GW emission dominates their evolution (equation~\ref{eq_Egw}). Hence, in equation~\ref{eqn:GWBlowerz} $\langle t_{\rm evol} \rangle$ appears as a parameter in front of the integral. At $z<0.7$ the integral turns into a summation over 88 objects with individual redshifts and mass distribution described in \S\ref{sec:massDistribution}. At $z \geq 0.7$ we integrate over the mass and redshift distribution of the SDSS quasars shown in Figure~\ref{fig:Qcatalog}, and normalize it relative to the number of SMBHB candidates at $z<0.7$.

\begin{figure*}[t]
\centering
\includegraphics[width=0.8\textwidth, clip=true]{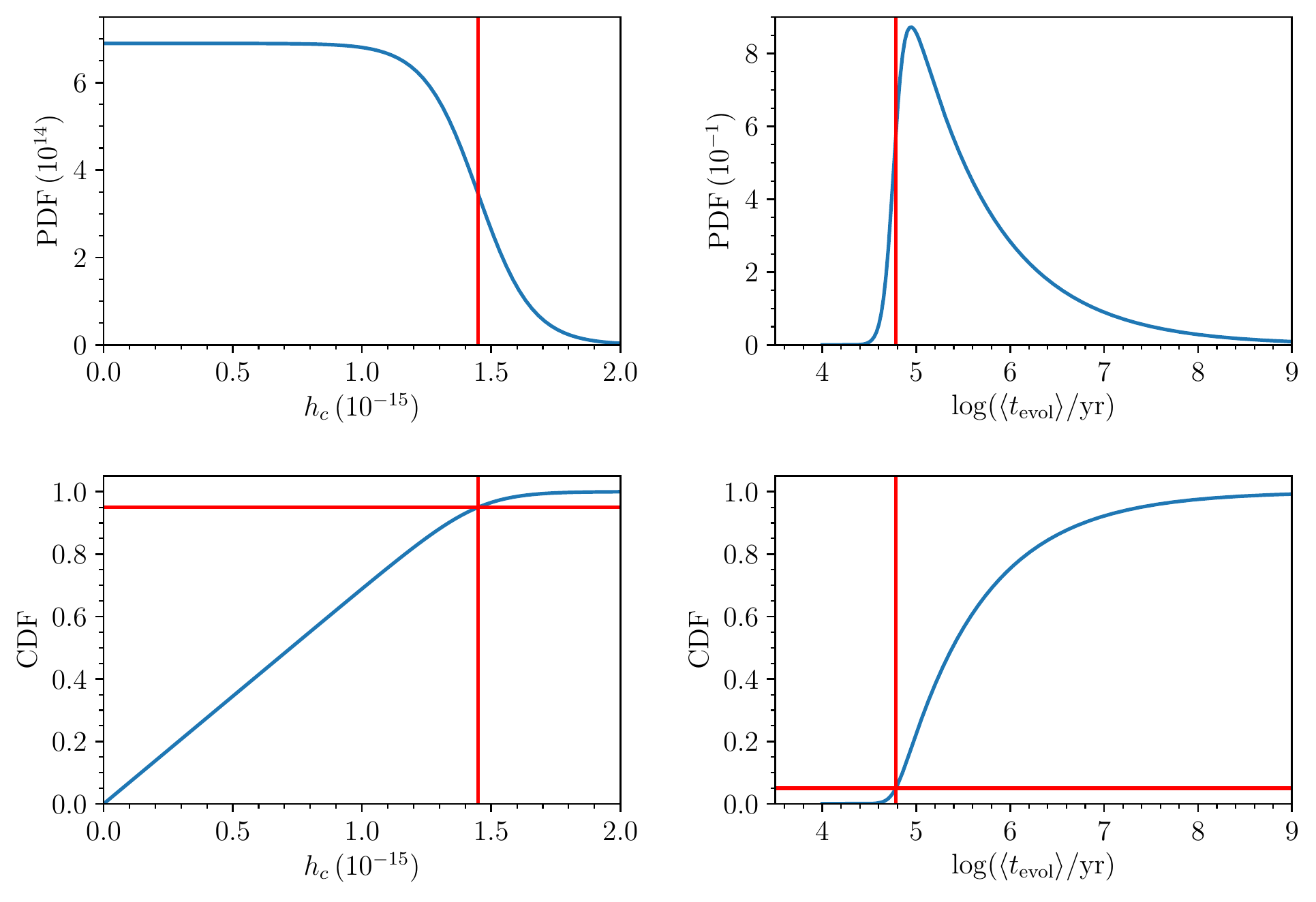}
\caption{{\it Top row:} PDFs for the GW strain amplitude, contributed by a population of inspiraling SMBHBs at a frequency $f= 1\, \textrm{yr}^{-1}$ ($h_c$; left) and the average merger time for the same population ($\langle t_{\rm evol} \rangle$; right). Both refer to the model $h_{c3}$, in which SMBHB masses are corrected for Malmquist bias. {\it Bottom row:} CDFs corresponding to the PDFs in the top row. Red lines mark the 95 and 5 percentile values of $h_c$ and $\langle t_{\rm evol} \rangle$, respectively.}
\label{fig:PDF}
\end{figure*}


\section{NANOGrav constraints on the merger timescale}\label{sec:results}

We use equation~\ref{eqn:GWBlowerz} to calculate the GWB strain from the population of putative SMBHBs inferred from the E12 sample given a merger timescale, $\langle t_{\rm evol} \rangle$. Specifically, we calculate the strain at a reference frequency $f= 1\, \textrm{yr}^{-1}$ and summarize the results in Table~\ref{table:hc}. The first column of the table shows the value of $\langle t_{\rm evol} \rangle$ and the second shows the corresponding GWB strain, $h_{c1}$, calculated for a population of SMBHBs in the redshift range $0 < z < 0.7$, equivalent to that of the E12 sample. In this scenario the mass distribution of SMBHBs was not corrected for Malmquist bias. Note that for all values $h_c^2 \propto 1/\langle t_{\rm evol}\rangle $, so the table illustrates how different evolution times of SMBHBs affect the resulting amplitude of the GWB strain. Namely, longer $\langle t_{\rm evol} \rangle$ implies slower inspiral of binaries from subparsec scales to coalescence, and consequently, lower GWB.  

The third column of Table~\ref{table:hc} shows the strain amplitude, $h_{c2}$, calculated for a population of SMBHBs in the full redshift range, $0 < z < 1.5$, with masses uncorrected for Malmquist bias. Comparison of models $h_{c1}$ and $h_{c2}$ shows that when the contribution to the GWB from binaries with $z \geq 0.7$ is included, the overall strain amplitude increases by about 16\%. Low redshift SMBHBs therefore dominate the stochastic GWB at $f= 1\, \textrm{yr}^{-1}$ by a large margin. Hence, even if there is a population of low luminosity or higher redshift SMBHBs, not captured by the flux-limited SDSS spectroscopic survey, their contribution to the GWB should be small.

The fourth column shows $h_{c3}$, calculated for a population of SMBHBs with $0 < z < 1.5$ with masses corrected for Malmquist bias. Because the corrected mass distribution is characterized by a lower median value, in this case the overall GWB amplitude decreases by a factor of approximately 3 relative to $h_{c2}$. 

In the next step, we compare the calculated strain amplitudes in Table~\ref{table:hc} to the latest constraints provided by the 11\,yr NANOGrav data set, which sets a 95\% upper limit on the GW strain amplitude of $A_{\rm GWB}  < 1.45\times10^{-15}$ for SMBHBs emitting at a frequency of $1\,{\rm yr}^{-1}$  \citep{arzoumanian18}. Although this limit is a factor $\sim 1.5$ less stringent than that published by \citet{shannon15}, it includes a self-consistent Bayesian model of the solar system ephemeris, making it more robust. 

The top left panel of Figure~\ref{fig:PDF} shows the probability density function (PDF), corresponding to the model corrected for Malmquist bias ($h_{c3}$), which specifies the probability that $A_{\rm GWB}$ falls within a particular range of values. It is commonly modeled by a Fermi-like function \citep[e.g.,][]{chen17} 
\begin{equation}
PDF(h_c) = \frac{C_1}{1+\exp\left(\frac{h_c-A_{95}}{C_2}\right)} \,,
\end{equation} 
where $C_1=6.90\times 10^{14}$ and $C_2=1.05\times 10^{-16}$ are constants determined from PDF normalization and a requirement that 95 percentile value of the strain amplitude is $A_{95} = 1.45\times10^{-15}$, respectively.  The bottom left panel of Figure~\ref{fig:PDF}, shows the resulting cumulative distribution function (CDF), which indicates the probability that $A_{\rm GWB}$ is less than or equal to a given strain amplitude shown on the $x$-axis. The vertical line marks $A_{95}$, which corresponds to the sensitivity limit of NANOGrav at $f = 1\,{\rm yr}^{-1}$ \citep{arzoumanian18}.

The top right panel of Figure~\ref{fig:PDF} shows a PDF for model $h_{c3}$, of the merger time corresponding to a given value of $h_c$, such that ${\rm PDF}(\langle t_{\rm evol} \rangle)={\rm PDF}(h_c) |dh_c/d\langle t_{\rm evol}\rangle |$. The inferred distribution for $\langle t_{\rm evol} \rangle$ peaks at about $10^5\,$yr and indicates that there cannot be many subparsec binaries that evolve to merger on timescales $\ll 10^5\,$yr, since they would produce strain amplitudes $h_c \gg A_{95}$, and would already be detected by NANOGrav. Similarly, low strain amplitudes ($h_c \ll A_{95}$) can be produced by a small number of relatively slowly evolving binaries with $\langle t_{\rm evol} \rangle > 10^7\,$yr, illustrated by the extended tail of the distribution. 

Similarly to ${\rm PDF}(h_c)$, which provides an upper limit on the GWB strain created by inspiraling SMBHBs, ${\rm PDF}(\langle t_{\rm evol} \rangle)$ can be used to infer a lower limit on $\langle t_{\rm evol} \rangle$ for the same population of binaries. The lower right panel of Figure~\ref{fig:PDF} shows the CDF for $\langle t_{\rm evol} \rangle$ for model $h_{c3}$ and indicates that 95\% of the SMBHBs would have to evolve on timescales $\langle t_{\rm evol} \rangle > 6\times 10^4\,$yr in order to be consistent with the sensitivity limit of NANOGrav.  In comparison, the model where the SMBH mass distribution was not corrected for Malmquist bias ($h_{c2}$) predicts the peak of the distribution at about $8\times10^5\,$yr and $\langle t_{\rm evol} \rangle > 5\times 10^5\,$yr for 95\% of the SMBHBs.


\section{Discussion and Conclusions}\label{sec:discussion}

In this Letter we consider the contribution to the strain of a stochastic GWB from an expected population of inspiraling SMBHBs with redshift $z<1.5$, inferred from a sample of 88 subparsec SMBHB candidates discovered by the E12 spectroscopic search. We find that the average timescale for evolution of such SMBHBs from subparsec separations to coalescence must be $\langle t_{\rm evol} \rangle > 6\times 10^4\,$yr in order for the amplitude of their GWB to be consistent with the upper limit placed by NANOGrav. This limit is in agreement with a range of timescales ($\sim 10^6-10^9$\,yr) predicted by theoretical models for SMBHBs of similar properties, that evolve due to interactions with stars and / or gas in their host galaxies, and eventually merge due to the emission of GWs \citep[e.g.,][]{lodato09, haiman09,rafikov13}. This implies that, based on this test alone and within the uncertainties of theoretical models, all 88 SMBHB candidates from the E12 sample are presently consistent with being true binaries. It is of course plausible that only a fraction (or none) of the E12 candidates are actual SMBHBs -- if so, $\langle t_{\rm evol} \rangle$ would be reduced proportionally. Our results are subject to several assumptions which we discuss below.

\begin{itemize}
\item In this work we consider a population of hypothetical subparsec SMBHBs that appear as luminous SDSS quasars but do not account for the presence of SMBHBs in inactive galaxies. If SMBHBs in inactive galaxies are common, they could contribute to the stochastic GWB even if they are not found by the electromagnetic searches. For example, if the frequency of SMBHBs in inactive galaxies is similar to that in AGNs, then the underlying population of binaries could be $\sim 10$ times larger than the number inferred from the EM searches. If so, this would imply $\sim 10$ times longer merger timescale, $\langle t_{\rm evol} \rangle > 6\times 10^5\,$yr.

\item We assume that the mass distribution of primary SMBHs in binaries that contribute to the GWB is the same as that of the SMBHs that power SDSS quasars. This approach allows us to sidestep complications related to single-epoch virial mass measurements in potential SMBHBs, as those methods may not be applicable to binaries. Even so, the distribution of virial SMBH masses adopted in this work is subject to Malmquist bias, which shifts the distribution of measured masses to higher values by a factor of about 3 relative to the true underlying distribution. We find that if correction for this effect is omitted, the resulting merger timescale is $\sim 8$ times longer ($\langle t_{\rm evol} \rangle > 5\times 10^5\,$yr) than that for the scenario where this correction is applied. This example illustrates that somewhat different assumptions about the SMBHB mass function can lead to uncertainties of one order of magnitude in the limit on $\langle t_{\rm evol} \rangle$, and still be consistent with the upper limit on stochastic GWB. 

\item Another assumption we adopt is that the radial velocities measured in spectroscopic searches for SMBHBs trace the motion of the primary SMBHs. If instead the motion traced is that of the secondary, it is more natural to assume that the mass distribution of the secondary (as opposed to the primary) SMBHs is represented by that of the SDSS quasars. Because binaries with higher mass ratios are favored, the total mass of such systems would be similar (within a factor of 2) to the case when the primary's motion is traced. This results in $\langle t_{\rm evol} \rangle$ that is also within a factor of 2 of the value calculated for that scenario. Thus, we do not expect our results to be very sensitive to the assumption that spectroscopic searches trace the motion of the primary SMBHs.

\item An important assumption of this work is that SMBHBs that contribute to the GWB in the frequency band of NANOGrav inspiral only due to the emission of GWs. For example, this implies that the evolution of $\sim 10^8\,M_{\odot}$ SMBHBs with comparable mass ratios is dominated by GW emission when they reach separations of $\sim {\rm few}\times 10^{-3}\,$pc. While this assumption is justified for some binaries, the possibility that the evolution of SMBHBs at these separations is driven by gas or stellar torques cannot be eliminated for all. If so, such SMBHBs would evolve faster through the PTA frequency band, emitting with a lower strain amplitude relative to the scenario in which GW emission dominates. Therefore, the presence of additional physical mechanisms results in a lower, more conservative lower limit $\langle t_{\rm evol} \rangle$ than that based on the GW emission alone. Along similar lines, gas and stellar torques can in principle excite eccentricity of the SMBHB orbits, in which case our assumption of circular binaries would need to be revised.
\end{itemize}

In summary, this work illustrates an important place occupied by PTAs and observatories  that can provide independent tests of the nature of SMBHBs. While subparsec SMBHBs are still challenging to unambiguously identify, constraints like the one presented here keep narrowing down the range of possibilities for these objects.

\acknowledgements
S.R.T. acknowledges ongoing discussions with the NANOGrav and International Pulsar Timing Array collaborations. T.B. acknowledges the support by the National Aeronautics and Space Administration (NASA) under award No. 80NSSC19K0319 and by the National Science Foundation (NSF) under award No. 1908042.  T.B. and A.S. acknowledge partial support by the National Science Foundation under Grant No. NSF PHY-1748958 during their visit to the Kavli Institute for Theoretical Physics, where an idea for this work was conceived.  A.S. is supported by the European Research Council (ERC) under the European Union's Horizon 2020 research and innovation program ERC-2018-COG under grant agreement No 818691 (B Massive).



\bibliographystyle{aasjournal}
\bibliography{apj-jour,smbh}

\end{document}